\documentclass[12pt]{article}

\def\D{\Delta}
\def\d{\delta}

\def\l{\lambda}

\def\G{\Gamma}
\def\g{\gamma}
\def\e{\epsilon}
\def\s{\sigma}
\def\o{\omega}

\def\b{\beta}
\def\m{\mu}

\def\s{\sigma}

\def\e{\epsilon}

\def\det{\textrm{det}}

\newcommand{\be}{\begin{equation}}
\newcommand{\ee}{\end{equation}}
\newcommand{\bea}{\begin{eqnarray}}
\newcommand{\eea}{\end{eqnarray}}

\begin{document}

\begin{center}
\bf{\large  SPIN-CUBE MODELS OF QUANTUM GRAVITY}
\end{center}

\bigskip
\begin{center}
A. MIKOVI\'C \\
Departamento de Matem\'atica  \\
Universidade Lus\'ofona de Humanidades e Tecnologias\\
Av. do Campo Grande, 376, 1749-024 Lisboa, Portugal\\
and\\
Grupo de Fisica Matem\'atica da Universidade de Lisboa\\
Av. Prof. Gama Pinto, 2, 1649-003 Lisboa, Portugal\\
\end{center}

\centerline{E-mail: amikovic@ulusofona.pt}

\bigskip
\bigskip
\begin{quotation}
\noindent\small{We study the state-sum models of quantum gravity based on a representation 2-category of the Poincare 2-group. We call them spin-cube models, since they are categorical generalizations of spin-foam models. A spin-cube state sum can be considered as a path integral for a constrained 2-BF theory, and depending on how the constraints are imposed, a spin-cube state sum can be reduced to a path integral for the area-Regge model with the edge-length constraints, or to a path integral for the Regge model. We also show that the effective actions for these spin-cube models have the correct classical limit.}\end{quotation}

\bigskip
\bigskip
\noindent{\bf{1. Introduction}}

\bigskip
\noindent Spin foam models are discrete path-integral formulations of gauge theories and quantum gravity, see \cite{ba,ro}. The path integral for a spin foam model is defined as a state sum for a colored dual 2-complex of the spacetime manifold triangulation and the colors are choosen to be the objects and the morphisms of a representation category of the relevant symmetry group. In the case of General Relativity (GR) this group is the Lorentz group. A natural categorical generalization of a spin foam model would be a state sum model based on a colored 3-complex, where the colors are objects, morphisms and 2-morphisms of a 2-category representation of the relevant 2-group, see \cite{bnw,mv2p}. We will refere to these models as spin cube models, and in the case of GR the relevant 2-groups are the Poincare 2-group \cite{mv2p} and the teleparallel 2-group \cite{bzw}. 

If one labels the 3-cells, 2-cells and 1-cells of a given 3-complex with the objects, morphisms and 2-morphisms of a given 2-category, this is equivalent to labelling the edges, triangles and tetrahedrons of a spacetime triangulation. Hence the spin cube models give a possibility of introducing the edge lengths as deegres of freedom, beside the triangle spins and the tetrahedron intertwiners, which are the spin foam variables. In the case of the Poincare 2-group there is a representation 2-category such that the objects (representations) are labelled by positive numbers. These representations satisfy the triangle inequalities when composed and the corresponding intertwiners are $U(1)$ spins for non-zero area triangles \cite{cs,bbfw}. 

The reason why one would like to introduce the edge lengths as additional degrees of freedom, is that in this way one can solve the problems of spin foam models related with the fact that an arbitrary spin-foam configuration does  not correspond to a metric geometry. Namely, the spins of triangles in a spin foam model correspond to the areas of triangles, and an arbitrary assignement of triangle areas does not give a well-defined metric geometry \cite{brw, ww,fs}, unless the edge-length constraints are imposed \cite{mak}. In the current formulations of spin foam models \cite{eprl,fk}, there are no Lagrange multipliers which would impose the edge-length constraints and therefore the only possibility for these constraints to appear is dynamically, which is not guaranteed and it is difficult to verify.

Consequently, it is difficult to couple fermionic matter to spin foam models, since the fermions couple to the edge lengths, and these are not well defined in an arbitrary spin foam configuration. Also, when the effective action is computed in the semi-classical approximation, the classical limit is the area-Regge action \cite{mvea,mvea2}. Hence  the classical limit for smooth spacetimes can not be automatically identified with the Einstein-Hilbert action.  Although there are indications that the edge-length constraints may appear dynamically \cite{mvea}, it is difficult to prove that the usual Regge action will appear. The presence of the edge-length variables in spin cube models solves automatically the problem of coupling of fermionic matter, while the effective action for a spin cube model can naturally have the usual Regge action as its classical limit. 

The study of spin cube models started in \cite{mv2p}, and there it was argued that a topological spin cube state sum can be transformed into a quantum gravity one by imposing the constraints which relate a triangle spin to the area of the triangle. Since the relationship between the triangle spin and the triangle area is not unique, in this paper we will show that it is possible to implement the GR constraints such that the independent variables are the edge lengths. In this case the spin-cube weights can be choosen such that the state sum reduces to the Regge model path integral for GR. We also show that it is possible to implement the GR constraints such that the triangle spins are left as the independent variables, in which case the state sum reduces to a path integral for the area-Regge model with the edge-length constraints.

In section 2 we review breifly the Poincare 2-group and its relationship with GR. We also review the construction of a state sum for a Poincare 2-group representation 2-category, which is relevant for quantum gravity. In section 3 we discuss the implementation of the GR constraints on the spin cube state sum, and we show how to implement them such that a solution in terms of the triangle spins is obtained. This solution gives a spin foam model which is a discretization of a path integral for the area-Regge model with the edge-length constraints. A slight modification of the spin-cube weights gives a spin foam model such that one can easily show that the classical limit of the effective action is the area-Regge action with the edge-length constraints. In section 4 we implement the GR constraints in the state sum such that the independent variables are the edge lengths, and the state sum becomes a discretized path integral for the Regge model. By using the effective action technique, we show that the classical limit is the Regge action. In section 5 we present our conclussions. 

\bigskip
\bigskip
\noindent{\bf{2. Poincare 2-group state sum models}}

\bigskip
\noindent A 2-group is a categorification of a group, since a group is an invertible category with one object, while a 2-group is an invertible 2-category with one object, see \cite{bh}. Any 2-group is equivalent to a crossed module, and the latter is simply a pair of groups $G$ and $H$ such that there is a map $\partial : H \to G$ which is a homomorphism and a map $\triangleright : G \times H \to H$, which is a group action, such that 
$$\partial (g \triangleright h ) = g (\partial h ) g^{-1} \,,\quad (\partial h) \triangleright h' = h h' h^{-1} \,,$$ 
where $g \in G$ and $h,h' \in H$.

A tipical example is the $n$-dimensional Euclidean 2-group, where $G=SO(n)$ and $H={\bf R}^n$. The $\partial$ map is trivial while the $\triangleright$ map is the usual action of a rotation on a vector.  The semi-direct product $G \times_s H$ corresponds to the group of 2-morphisms in a 2-group, so that the usual Poincare group is only a part of the Poincare 2-group where $G=SO(3,1)$ and $H={\bf R}^4$.

The reason why the Poincare 2-group is relevant for GR is that GR can be represented as a gauge theory for the Poincare 2-group \cite{mv2p}. More precisely, the Einstein equations can be derived from an action which describes a constrained 2-BF theory for the Poincare 2-group
\be S = \int_M \left[B^{ab} \wedge R_{ab} + e^a \wedge \nabla\beta_a - \l^{ab} (B_{ab} - \e_{abcd}\,e^c\wedge e^d)\right] \,,\label{2pgra}\ee
where $R_{ab}$ is the curvature 2-form for the Lorentz group connection $\o_{ab}$ and $\b_a$ is a 2-form which together with $\o_{ab}$ forms a 2-connection $(\o_{ab} , \b_a)$ for the Poincare 2-group. The 2-forms $B_{ab}$ and the one-forms $e_a$, which can be identified with the tetrads, enforce the vanishing of the 2-curvature
$$ (R_{ab}, \nabla \b_a ) = (d\o_{ab} + \o_{ac} \wedge \o^c_b, d\b_a + \o_{ab} \wedge \b^b )\,, $$
in the topological case, when $\lambda_{ab} = 0$. The constraint 
\be B_{ab} = \e_{abcd}\,e^c\wedge e^d \,, \label{grcs}\ee
transforms the topological gravity theory
$$S_{top} = \int_M (B^{ab}\wedge R_{ab} + e^a \wedge \nabla\b_a) \,,$$ 
into GR and it is the same constraint which is used in the case of spin foam models. However, in the Poincare 2-group case the GR constraint can be written in a simpler way since the tetrads appear explicitely in the theory.

A quantum gravity theory can be constructed by using the path integral based on the action (\ref{2pgra}), see 
\cite{mv2p}. This theory takes a form of a state-sum model for a colored dual 3-complex of a triangulation of the space-time manifold. The set of colors consists of positive numbers for the edges, which satisfy the triangle inequalities, while the colors for the triangles and the tetrahedrons can be the irreps and the corresponding intertwiners for the Lorentz group or its $SO(3)$ and $SO(2)$ subgroups. 

This result agrees with the categorical structure of a state sum for a 2-group, since the labels for the edges can be interpreted as the labels for 2-group representations, while the labels for the triangles can be interpreted as the corresponding intertwiners. The labels for the tetrahedrons can be interpreted as the 2-intertwiners, and they arise because a 2-group representation category is a 2-category, and hence the 2-intertwiners correspond to 2-morphisms.

In the Poincare/Euclidean 2-group case there is a 2-Hilbert space representation 2-category, see \cite{cs,bbfw}, such that the object (representation) labels are positive numbers. The corresponding triangle intertwiners are $SO(2)$ or $U(1)$ irreps if  the triangles have non-zero areas. The 2-intertwiner labels for the tetrahedra are trivial, so that one can construct a state sum as
\be Z = \int_{\tilde{\bf R}_+^{E}} \prod_{\e=1}^E \m (L_\e)\, dL_\e \sum_{m\in {\bf Z}^{F}} \prod_{\D=1}^F W_\D (L,m) \prod_{\s=1}^V W_\s (L,m) \,,\label{gscss}\ee
where $\e$ are the edges of a triangulation $T(M)$ of the 4-manifold $M$, $\D$ are the triangles of $T(M)$ and $\s$ are the 4-simplices of $T(M)$. $E$ is the number of edges, $F$ is the number of triangles, $V$ is the number of 4-simplices and $\tilde{\bf R}_+^E$ is the subset of ${\bf R}_+^E$ whose elements satisfy the triangle inequalities associated with the triangulation $T(M)$.

The weights $\m_\e$, $W_\D$ and $W_\s$ should be chosen such that the state sum $Z$ resembles a discretized path integral for GR. More precisely, a choice of the weights should be such that it implements the GR constraint (\ref{grcs}) and that the corresponding state-sum model defines a quantum gravity theory whose classical limit is the
Regge action
\be S_R  = \sum_{\D=1}^F A_\D (L)\, \theta_\D (L) \,,\label{lr}\ee
where $A_\D$ is the area of a triangle $\D$ and $\theta_\D$ is the deficit angle. We will refere to (\ref{lr}) as the length-Regge action in order to distinguish it from the area-Regge action
\be S_{AR} = \sum_{\D=1}^F A_\D \, \theta_\D (A) \,,\label{ar}\ee 
which can be naturally associated to a spin foam model.

\bigskip
\bigskip
\noindent{\bf{3. State sum with the GR constraint}}

\bigskip
\noindent The GR constraint (\ref{grcs}) can take the following form in the discrete setting
\be \g m_\D = A_\D (L) \,, \label{dc}\ee
where $m_\D \in {\bf N}$ is an $SO(2)$ spin of a triangle $\D$, $A_\D (L)$ is the area of a triangle with edge lengths $L_1,L_2$ and $L_3$ and $\g$ is a constant, which is analogous to the Barbero-Immirzi constant which appears in the case of spin foam models. In order to have simpler formulas, we are going to take $\g =1$. The function $A(L)$ is given by Heron's formula
\be A(L) = \sqrt{s (s-L_1) (s-L_2) (s-L_3)} \,,\ee
where $2s = L_1 + L_2 + L_3$ is the triangle perimeter. 

In order to get physical lengths and areas one has to make the rescaling $L \to L / l_0$ in (\ref{dc}), where $l_0$ is a unit of length. It is natural to choose $l_0$ to be the Planck length $l_P$. Note that choosing $l_0$ to be a multiple of $l_P$ is equivalent to choosing $\g \ne 1$.

The constraints (\ref{dc}) can be implemented in the state sum (\ref{gscss}) by choosing the triangle weights as
\be  W_\D =  \d ( m_\D - A_\D (L) ) \,.\label{sgrc}\ee
In order to insure that the Regge action will be the classical limit of the model, we will choose
\be W_\s =\exp \left( i \sum_{\Delta\in\sigma}  m_\Delta \,\theta^{(\s)}_\Delta (L)\right) \,, \label{rw}\ee
where $\theta^{(\s)}_\D (L)$ is the interior dihedral angle \cite{mv2p}. The reason for this choice is simple to understand, since
$$ \prod_{\sigma=1}^V \exp \left( i \sum_{\Delta\in\sigma}  m_\Delta \,\theta^{(\s)}_\Delta (L)\right) = \prod_{\sigma=1}^V \exp \left( i \sum_{\Delta\in\sigma} A_\Delta (L) \,\theta^{(\s)}_\Delta (L)\right) \,, $$
due to the constraint $ m_\D = A_\D (L)$, so that
$$ \prod_{\sigma=1}^V \exp \left( i \sum_{\Delta\in\sigma} A_\Delta (L) \,\theta^{(\s)}_\Delta (L)\right) = e^{iS_R (L)}\,.$$

Hence the constraints (\ref{dc}) can reduce the spin-cube state sum to a path integral for the Regge model. However, there are certain caveats in this simple reasoning, which we will demonstrate by an exact analysis. Let us start from the state sum with the weights (\ref{sgrc}) and (\ref{rw})
\be Z = \sum_{m\in {\bf N}^{F}}\, \int_{\tilde{\bf R}_+^{E}} \, \prod_{\epsilon=1}^E \m (L_\e )\, dL_\epsilon\, \prod_{\Delta=1}^F \delta ( m_\Delta - A_\Delta (L))\,\prod_{\sigma=1}^V \exp \left( i \sum_{\Delta\in\sigma} m_\Delta \,\theta^{(\s)}_\Delta (L)\right) \,. \label{grz}\ee

The form of (\ref{grz}) suggests to integrate first the lengths, which will transform (\ref{grz}) into a sum over the spins subject to the constraints
\be m_f - A_f (L) = 0 \,,\quad f = 1,2,...,F \,.\label{grc}\ee
In order to solve these constraints, note that in a four-manifold triangulation we have 
$$F \ge \frac{4}{3}E \,, $$ 
since $F$ triangles have $3F$ edges, and each edge is shared by at least $4$ triangles, so that $3F \ge 4E$. Consequently 
$$F > E \,,$$ 
so that we can solve the first $E$ constraints of (\ref{grc}) as 
\be L_\e = l_\e (m_1 ,\cdots ,m_E ) \,,\ee 
where $\e = 1,2,..., E$, while the remaining $F - E$ constraints become the Diofantine equations 
\be m_k = \varphi_k (m_1,...,m_E) \,,\quad E+1 \le k \le F \,, \label{diof}\ee 
where $\varphi_k (m) = A_k (l(m))$. Hence $m\in D_F \subset {\bf N}^{F}$. However, it is difficult to determine the structure of $D_F$ and it may be the empty set.

This problem can be solved by relaxing the constraints (\ref{diof}) as
\be m_k = [\varphi_k (m_1,...,m_E)] \,,\quad E+1 \le k \le F \,, \label{lc}\ee 
where $[x]$ is the integer part of a real number $x$. In this case the constraints are given by 
$$ m_e =  A_e (L) \,,  \quad 1 \le e \le E \,,$$
\be m_k = [ A_k (L)] \,, \quad E+1 \le k \le F \,,\label{csone}\ee
and the solution is $L_\e = l_\e (m')$ where $m' \in {\bf N}^E$ and $m'' = [\varphi (m')] \in {\bf N}^{F-E}$.
Since the functions $l_\e (m')$ have to be real, this means that $m' \in D_E \subset {\bf N}^E$, which is related to the fact that $L_\e$ have to satisfy the triangle inequalities.

Let us now introduce the new weights in the spin-cube state sum, so that we start from (\ref{gscss}) with
\be \prod_{\D=1}^F W_\D (L,m) = \prod_{f=1}^E \delta (m_f - A_f (L)) \prod_{f=E+1}^F \delta (m_f - [A_f (L)]) \label{sfw}\ee
and $W_\s$ is given by (\ref{rw}).
By integrating the $L$ variables we obtain the following spin foam model
\bea Z &=& \sum_{m\in D_E} \prod_{\e = 1}^E \m_\e (l(m))\,J(m_1,...,m_E)\, \exp \Big{(} i\sum_{f=1}^E m_f \theta_f (m) \cr &+& i\sum_{f=E+1}^F[\varphi_f (m)] \theta_f (m)
\Big{)} \,,\label{2psf}\eea
where
$$J(m_1,...,m_E) = \left|\frac{\partial (L_1,...,L_E)}{\partial (m_1,...,m_E)}\right| $$ 
is the Jacobian for $L_\e = l_\e (m)$.

Note that this is a spin foam model with a nonlocal weight 
\be W_E (m) = \prod_{\e = 1}^E \m_\e (l(m))\,J(m_1,...,m_E) \label{nlw} \ee
and the state sum has a form of a path integral for an area-Regge model
$$ Z = \sum_{m\in D_E} W_E (m) \exp\left(iS^*_{AR}(m)\right),$$ 
where 
$$ S^*_{AR}(m)= \sum_{f=1}^E m_f \theta_f (m) + \sum_{f=E+1}^F[\varphi_f (m)] \theta_f (m)\,.$$
This is an area-Regge action, with integer areas, where the edge-length constraints are imposed via (\ref{lc}). 

The finiteness and the effective action for the spin-foam model (\ref{2psf}) can be studied by using the techniques of \cite{mvea,mvea2,mvsff}. We will not do this here, since the analysis gets complicated due to the presence of the non-local weight (\ref{nlw}). 

Note that one can define a new model by choosing $\m (L_\e ) = 1$, $W_\s$ as in (\ref{rw}) and a non-local weight for the triangles in the spin-cube state sum
$$ \tilde W (L,m) =J^{-1}(m_1,\cdots,m_E ) \prod_{\D=1}^F W_\D (L,m) \prod_{\D=1}^E m_\D^{-p} \,, $$
where $W_\D$ are given by (\ref{sfw}). 
This choice of the weights gives a spin foam state sum model with local weights for the triangles
\be \tilde Z = \sum_{m\in D_E} \prod_{f = 1}^E m_f^{-p} \exp \left( i S^*_{AR}(m) \right) \,.\label{2par}\ee

The semiclassical effective action for the area-Regge spin foam model (\ref{2par}) can be easilly calculated  by using the results of  \cite{mvea,mvea2}. We obtain for $m\to \infty^E$
\be \G (m) = S^*_{AR}(m) + p\sum_{f=1}^E \ln m_f  + \frac{1}{2}\,Tr\left(\log (S^*_{AR})''(m)\right) + O(m^{-2})\,,\label{lwsfa}\ee
where $(S^*_{AR})''(m)$ is the hessian matrix for the function $(S^*_{AR})(m)$. Since
\be S^*_{AR}(m) = O(m) \,,\quad p\sum_{f=1}^E \ln m_f = O(\ln m) \,,\quad Tr\left(\log (S^*_{AR})''(m)\right)= O(m^{-1}) \,,\label{eata}\ee
where the notation $f(m) = O(m^r)$ means that
$$ f(\l m_1 ,\cdots, \l m_E ) \approx \l^r g(m,\l) $$
and $f(m) = O(\ln m)$ means
$$ f(\l m_1 ,\cdots, \l m_E ) \approx (\ln\l )\, g(m,\l) $$
for $\l\to\infty$ and $g(m,\l)$ is a bounded function of $\l$. From (\ref{eata}) it follows that the classical limit of the effective action (\ref{lwsfa}) will be the area-Regge action $S^*_{AR}(m)$. However, the action $S^*_{AR}(m)$ is dynamically equivalent to the length-Regge action $S_R (L)$ due to the constraints (\ref{lc}). 

As far as the convergence of the state sum (\ref{2par}) is concerned, it is easy to see that it is absolutely convergent for $p>1$, while the convergence for $p\le 1$ case is a more complicated issue and will not be analysed here. 

\bigskip
\bigskip
\noindent{\bf{4. Edge-length state sum models}}

\bigskip
\noindent The spin foam model (\ref{2psf}) appeared because we integrated the edge-lengths first in the spin cube state sum. This was a natural way to proceed, because of the delta-function weights (\ref{sgrc}) and the fact that the spins $m$ are integers. A natural question to ask is it possible to implement the constraints such that the edge lengths remain as the independent variables.

A clue comes from the relaxed constraints (\ref{csone}), so that let us consider the following set of constraints
\be m_f = [ A_f (L)]\,,\quad f = 1,2,..., F\,. \label{cstwo}\ee
These constraints have solutions for any $L \in \tilde{\bf R}_+^E$, and if we take
$$ W_f (L,m) = \delta (m_f - [A_f (L)])\,,$$
with $W_\s$ given by (\ref{rw}), then the sumation over the spins $m$ in (\ref{gscss}) gives
\be Z = \int_{ \tilde{\bf R}_+^{E}} \, \prod_{\e=1}^E \mu_\e (L) \, dL_\e \,\exp \left( i\tilde{S}_R (L)\right) \,, \label{rss}\ee
where
$$ \tilde{S}_R = \sum_{\D=1}^F  [ A_\D (L)] \theta_\D (L) \,. $$

Hence the constraints (\ref{cstwo}) reduce the state sum to a path integral for a continious-length integer-area Regge model. The mesure $\mu$ can be choosen such that $Z$ is finite. For example
\be \m (L_\e ) = (1 + L_\e )^{-p} \,,\label{regmu}\ee
will give an absolutely convergent partition function for $p>1$, since
$$ |Z| \le \int_{\tilde{\bf R}_+^E} \prod_{\e=1}^E (1 + L_\e )^{-p}\,dL_\e < \int_{ {\bf R}_+^{E}} \prod_{\e=1}^E (1 + L_\e )^{-p}\,dL_\e \,,$$
so that
\be |Z| < \left(\int_
{0}^{+\infty} \frac{dL}{(1 + L)^p}\right)^E \label{finz}\,.\ee
The integral in (\ref{finz}) is convergent for $p>1$. More generally, $\mu$ can be chosen such that $\m(0)$ is finite and $\m(L) = O(L^{-p})$ where $p \in {\bf R}$. However, the convergence of the state sum for $p \le 1$ case is a more complicated problem and we will not attempt to resolve it here.

The effective action $\G(L)$ can be found as a solution of the following integro-differential equation
\be e^{i\G (L)} = \int_{ {\bf R}_+^{E}} \, \prod_{\e=1}^E \mu (L_\e + l_\e)\,dl_\e \, \exp \left( i\tilde S_R (L+l)- i\sum_{\e=1}^E \frac{\partial\G}{\partial L_\e }\,l_\e\right) \,,\label{clr}\ee
see \cite{mvea2}. Note that the quantum fluctuations $l_\e$ do not satisfy the triangle inequalities so that the integration region is ${\bf R}_+^E$. This is a natural requirement, which is also reinforced by the fact that requiring the triangle inequalities for the quantum fluctuations would prevent obtaining closed-form results for the quantum corrections.

In the case when the background lengths are large ($L_\e >> 1$) the equation (\ref{clr}) can be solved perturbatively as
\be  \G(L) = \sum_{n\ge 0} \G_n (L) + \textrm{const} \,, \label{pex}\ee
where
$$ \G_0 (L) = \tilde S_R (L) -i \sum_{\e=1}^E \log \m (L_\e ) \,, $$
while
\be \G_n (L) = O(L^{-n+\nu(n)}) \,, \label{las}\ee
for $n \ge 1$, where $\nu(n) = \d_{n,1}$.

The explicit form of the perturbative terms $\G_n (L)$ can be obtained by introducing a formal perturbative parameter $\varepsilon$ such that
$$ \G(L,\varepsilon) = \sum_{n\ge 0} \varepsilon^n \G_n (L) +\textrm{const} $$
where $ \G(L,\varepsilon)$ is a solution of 
\be  e^{i\G /\varepsilon} = \int_{ {\bf R}_+^{E}} \, \prod_{\e=1}^E dl_\e \, \exp \left( \frac{i}{\varepsilon}S_\m (L+l)- \frac{i}{\varepsilon}\sum_{\e=1}^E \frac{\partial\G}{\partial L_\e }\,l_\e\right) \,.\label{eclr}\ee
Here 
$$S_\m (L) = \tilde S_R (L) - i\sum_{\e=1}^E \log \mu (L_\e )$$
and the initial condition is
$\G_0 = S_\m $.

By substituting the Taylor expansions for $\tilde S_R (L+l)$ and $\log \mu (L+l)$ into (\ref{eclr}), one obtains
\be \G_1 (L) =  \frac{i}{2} \, Tr\left(\log \hat{S}_R''(L)\right)  \,,\ee
where 
$$ (\hat{S}''_R )_{\e\e'} = (\tilde{S}_R'')_{\e\e'} - ip\frac{\d_{\e,\e'}}{L_\e^2} \,,$$
and we have taken that $\m(L) \approx L^{-p}$ for large $L$.

A perturbative solution of (\ref{clr}) of the type (\ref{pex}) exists because the coefficients in the Taylor expansion 
$$\tilde{S}_R (L+l) = \tilde{S}_R (L) + \langle \tilde S_R '(L), l \rangle  + \frac{1}{2} \langle \tilde S_R''(L) l,l\rangle + \cdots \,, $$
satisfy
\be \tilde{S}_R^{(n)}(L) = O( L^{2-n})  \,,\label{ras}\ee
due to the fact that 
$$\tilde{S}_R (L) = S_R (L) + \delta S_R (L) \,,$$ 
where
$$ \delta S_R = - \sum_{\D=1}^F \{A_\D (L)\}\theta_\D (L) \,,$$
and $\{x\} = x -[x]$ is the decimal part of a real number $x$.

The asymptotics (\ref{ras}) follows from the fact that $S_R(L)$ is a homogenious function of degree $2$ and $\d S_R (L)$ is a homogenious function of degree zero, while a partial derivative of a homogenious function is a homogenious function of the degree smaller by one. The choice of $\m (L)$ has to be such that it has the asymptotics
\be \mu(L) = O(L^{-p}) \,,\label{muas}\ee
which is dictated by the reqirement that the Regge action is the classical limit of the effective action and that the quantum corrections are small for large $L$, which will be shown in the next paragraph.

Since $\tilde S_R (L) = O(L^2)$ and $\log\m (L) = O(\log L)$ due to (\ref{muas}), the terms in the expansion (\ref{pex}) satisfiy
$$ |\G_n (L)| >> |\G_{n+1} (L)| \,,$$
for $n\ge 0$, as well as
$$ |S_R (L)| >> |\sum_{\e=1}^E \log\m(L_\e)| >> |\d S_R (L)| \,.$$
This implies that the classical limit of $\G$ is the Regge action $S_R$, i.e.
$$ \G(L) \approx S_R (L) $$
for $L\to \infty$. 

Note that the solution (\ref{pex}) is not a real function, while a physical $\G(L)$ has to be a real function. The same problem occurs in Quantum Field Theory, where it is solved by using the Wick rotation $iS \to -S_E$, where $S$ is the action while $S_E$ is the action in a Euclidean background metric. In our case the Wick rotation transforms the equation (\ref{clr}) into
\be e^{-\G (L)} = \int_{ {\bf R}_+^{E}} \, \prod_{\e=1}^E \mu (L_\e + l_\e)\,dl_\e \, \exp \left( -\tilde{S}_{ER} (L+l)+ \sum_{\e=1}^E \frac{\partial\G}{\partial L_\e }\,l_\e\right) \,,\label{wreq}\ee
which clearly allows for real solutions. However, the equation (\ref{wreq}) will have perturbative solutions only if $\tilde{S}_{ER}(L)$ is a positive function, which is not the case. The reason why the equation (\ref{clr}) has perturbative solutions, while the Wick rotated version (\ref{wreq}) does not, comes from the fact that $\int_{\bf R} e^{iax^2} dx $, $a\in \bf R$, is defined for any sign of $a$, while $\int_{\bf R} e^{-ax^2} dx $ is only defined for $a>0$.

Hence we are going to solve perturbatively the original equation (\ref{clr}), and a real effective action will be obtained by the following transformation
\be  \G \to Re\, \G + Im\, \G \,,\label{realg}\ee
which was introduced in \cite{mvea} in the case of spin foam models. The prescription (\ref{realg}) then gives for a physical solution
\be \G (L) =  S_R (L) + \sum_{\e=1}^E p \ln L_\e + \d S_R (L) + \frac{1}{2}\,Tr\left(\log S_R''(L)\right)  + O(L^{-2}) \,.\label{foea}\ee

In order to derive (\ref{foea}) the crucial identity was
$$ \int_{{\bf R}^n} d^n x \,e^{i\langle x, A x\rangle + \langle b, x\rangle } = \left(i\pi\right)^{n/2}(\det A )^{-1/2} \, e^{\langle b, A^{-1} b \rangle /4}\,,$$
which is a consequence of the Fresnel  integrals, i.e.
$$ \int_{-\infty}^\infty d x \,e^{ia x^2 } = \sqrt{\frac{i\pi}{ a}}\,.$$

\bigskip
\bigskip
\noindent{\bf{5. Conclusions}}

\bigskip
\noindent By imposing the GR constraints (\ref{dc}) on the 2-group state sum strongly via the delta-function weight (\ref{sgrc}) we obtained
a spin foam model where the spins $m$ are solutions of the Diofantine equation (\ref{diof}). Since the structure of the solution set is unknown and difficult to analyse, we relaxed the GR constraints to a form (\ref{csone}) and
obtained an area-Regge spin foam model  with the geometric areas. The geometric areas appear because the spins $m$ are constrained such that they correspond to an assigment of lengths to the edges of the triangulation. The corresponding state sum takes a form of a path integral for the area-Regge action with the edge-length constraints and a non-local weight for the triangles. We expect that the corresponding effective action will have the length-Regge action as its classical limit. It is possible to modify the weights in the spin-cube state sum such that one obtains a spin foam model with local triangle weights (\ref{2par}),
and it is easy to show that this model has the length-Regge action as its classical limit.

If the GR constraints are further relaxed, such that each triangle spin is equal to the integer part of the triangle area, then the space of solutions is given by all possible edge lengths for a given triangulation. The corresponding state sum is a path integral for the length-Regge action with integer areas. The effective action can be calculated in the semi-classical limit and the classical limit is the usual length-Regge action.

Note that in the case of quantum Regge calculus, the path integral is given by the state sum (\ref{rss}) where the integer-area Regge action $\tilde S_R$ is replaced by the usual Regge action $S_R$. Then the semiclassical expansion of the effective action is given by (\ref{foea}) but without the $\d S_R$ term. 

Therefore we have constructed examples of state sum models of quantum gravity whose effective actions have classical limit which is the Regge action. By refining the triangulation, the Regge action becomes the Einstein-Hilbert action, and therefore we have constructed state sum models whose effective actions have GR as the classical limit. An important issue to study is how the classical limit of a spin-cube model effective action is related to the usual definition of the classical limit
$$ Im\,\log\Psi(L_b) \approx S_0 (L_b ) \,,$$
for $L_b$ large, where $\Psi (L_b)$ is a wavefunction for a 3-boundary $b$ and  $S_0 (L_b )$ is a solution of the Hamilton-Jacobi equation in the Hamiltonian formulation of a spatialy discretized GR where a metric on $b$ is replaced by the edge lengths $L_b$ of a triangulation of $b$. The wavefunction $\Psi (L_b)$  is given by the spin cube state sum for a 4-manifold whose boundary is $b$ and the boundary edge lengths are given by $L_b$.

Note that the semi-classical effective action is defined for any $p$, independently of whether $Z$ is convergent or not. However, if we want to find a non-perturbative solution, then it is important that $Z$ is convergent, and hence it is important to know what happens in $p \le 1$ cases. One way to determine the non-perturbative solutions is to use a computer. Note that the numerical techniques which have been developed in the case of
Casual Dynamical Triangulations (CDT) models \cite{cdt}, may be usefull for such a task, since these models are related to our state sum models. Namely, insted of fixing a triangulation and summing over various edge-length assignments, in the case of CDT models one sums over different triangulations with fixed edge lengths.

\bigskip 
\bigskip
\noindent{\bf Acknowledgments}

\bigskip
\noindent This work has been partially supported by the FCT projects PTDC/MAT/099 880/2008 and
PEst-OE/MAT/UI0208/2011.

\end{document}